\input epsf
\input harvmac

\newcount\figno
\figno=0

\def\fig#1#2#3{
\par\begingroup\parindent=0pt\leftskip=1cm\rightskip=1cm\parindent=0pt
\baselineskip=11pt
\global\advance\figno by 1
\midinsert
\epsfxsize=#3
\centerline{\epsfbox{#2}}
\vskip 12pt
{\bf Fig. \the\figno:} #1\par
\endinsert\endgroup\par
}

\def\figlabel#1{\xdef#1{\the\figno}}
\def\encadremath#1{\vbox{\hrule\hbox{\vrule\kern8pt\vbox{\kern8pt
\hbox{$\displaystyle #1$}\kern8pt}
\kern8pt\vrule}\hrule}}

\overfullrule=0pt

%%%%%%%%%%%%%%% macros %%%%%%%%%%%%%%%%%%%%%%%
%

\def\bar{\overline}

\def\np#1#2#3{Nucl. Phys. {\bf B#1} (#2) #3}
\def\pl#1#2#3{Phys. Lett. {\bf #1B} (#2) #3}

\def\prd#1#2#3{Phys. Rev. {\bf D#1} (#2) #3}

\font\zfont = cmss10 %scaled \magstep1
\font\litfont = cmr6

\def\bigone{\hbox{1\kern -.23em {\rm l}}}
\def\ZZ{\hbox{\zfont Z\kern-.4emZ}}
\def\half{{\litfont {1 \over 2}}}

%
%%%%%% Sangmin's Macros
%

\def\munu{{\mu\nu}}
\def\Ap{\alpha'}
\def\dint#1#2{\int\limits_{#1}^{#2}}
\def\goto{\rightarrow}
\def\para{\parallel}
\def\bz{\bar{z}}
\def\bw{\bar{w}}
\def\brac#1{\langle #1 \rangle}
\def\calA{{\cal A}}

%%% (re)define the Title command %%%

\def\Title#1#2{\rightline{#1}
\ifx\answ\bigans\nopagenumbers\pageno0\vskip1in%
\baselineskip 15pt plus 1pt minus 1pt
\else%\special{papersize=11in,8.5in}%
\def\listrefs{\footatend\vskip 1in\immediate\closeout\rfile\writestoppt
\baselineskip=14pt\centerline{{\bf References}}\bigskip{\frenchspacing%
\parindent=20pt\escapechar=` \input
refs.tmp\vfill\eject}\nonfrenchspacing}
\pageno1\vskip.8in\fi \centerline{\titlefont #2}\vskip .5in}

%%% The front page %%%

\Title{
\vbox{\baselineskip12pt\hbox{\tt hep-th/yymmddd}
\hbox{IASSNS-HEP-97-59}\hbox{PUPT-1705}\hbox{SNUTP 97-070}
}}
{\vbox{\centerline{Absorption and Recoil of}
\bigskip\centerline{Fundamental String by D-String}}}

\centerline{Sangmin Lee\foot{sangmin@princeton.edu.}}
\smallskip
\centerline{\sl Joseph Henry Laboratories, Princeton University}
\centerline{\sl Princeton, NJ 08544, USA}
\smallskip
\smallskip
\centerline{Soo-Jong Rey\foot{sjrey@sns.ias.edu}}
\centerline{\sl School of Natural Sciences, Institute for Advanced Study}
\centerline{\sl Olden Lane, Princeton, NJ 08540, USA,}
\centerline{\sl Physics Department, Seoul National University, Seoul 151-742 KOREA}
\bigskip

\medskip

\noindent
Inclusive absorption cross section of fundamental IIB string to D-string 
is calculated perturbatively. The leading order result agrees
with estimate based on stringy Higgs mechanism  via Cremmer-Scherk 
coupling. It is argued that the subleading order
correction is dominated by purely planar diagrams in the large mass limit.
The correction represents 
conversion of binding energy into local recoil process of 
the fundamental string and D-string bound state. We show their presence 
explicitly in the next leading order. 
\Date{June, 1997}

%%%%%%%%%%%%%%%%%% References %%%%%%%%%%%%%%%%

\nref\polchinski
{J. Polchinski, \sl Dirichlet Branes and Ramond-Ramond Charges, 
\rm Phys. Rev. Lett. {\bf75} (1995) 4724;\ \
\sl TASI Lectures on D-brane, \tt hep-th/9611050. \rm}

\nref\sethi
{S. Sethi and M. Stern, \sl D-Brane Bound State Redux, 
\tt hep-th/9705046 \rm.}

\nref\schwarz{
J.H. Schwarz, \sl An SL(2,Z) Multiplet of Type II Superstrings, 
\rm \pl{360}{1995}{13}; \sl erratum \rm \pl{364}{1995}{252}.}

\nref\witten
{E. Witten, \sl Bound States of Strings and p-Branes, 
\rm \np{460}{1996}{335}.}

\nref\reyhiggs{
S.-J. Rey, \sl The Higgs Mechanism for Kalb-Ramond Gauge Field, 
\rm \prd{40}{1989}{3396}.}

\nref\cremmerscherk{
E. Cremmer and J. Scherk, 
\sl Spontaneous Dynamical Breaking of Gauge Symmetry in Dual Models, 
\rm \np{72}{1974}{117}.}

\nref\dai
{J. Dai and J. Polchinski, \sl The Decay of Macroscopic Fundamental Strings,
\rm \pl{220}{1989}{387}.}

\nref\marcus
{N. Marcus, \sl Unitarity and Regularized Divergences in String Amplitudes, 
\rm \pl{219}{1989}{265}.}

\nref\tsuchiya{
D. Mitchell, N. Turok, R. Wilkinson and P. Jeter,
 \sl The Decay Rate of Highly Excited Open String, \rm \np{315}{1989}{1}, 
\sl erratum \rm \np{322}{1990}{628};\ \
R.B. Wilkinson, N. Turok and D. Mitchell, 
\sl The Decay of Highly Excited Closed Strings, 
\rm \np{332}{1990}{131};\ \
K. Amano and A. Tsuchiya, \sl Mass Splittings and the Finiteness Problem
of Mass Shifts in the Type-II Superstring at One Loop, 
\rm \prd{39}{1989}{565};\ \
H. Okada and A. Tsuchiya, \sl Massive Modes in Type-I Superstring, 
\rm \pl{232}{1989}{91};\ \
D. Mitchell, B. Sundborg and N. Turok, \sl Decays of Massive Open Strings, 
\rm \np{335}{1990}{621};\ \
F. Lizzi and I. Senda, \sl Total Interaction Rate of Highly ExcitedStrings, 
\rm \pl{246}{1990}{385}.
}

\nref\myers
{M.R. Garousi and R. Myers, \sl Superstring Scattering from D-branes, 
\rm \np{475}{1996}{193}}

\nref\aki{
A. Hashimoto and I. Klebanov, \sl Decay of Excited D-branes, 
\rm \pl{381}{1996}{437};\ \
\sl Scattering of Strings from D-branes, \tt hep-th/9611214. \rm}

\nref\verlinde
{H. Verlinde, \sl A Matrix String Interpretation of the Large N Loop Equation, 
\rm UTFA-97/16, \tt hep-th/9705029 \rm.}

\nref\reyosaka{
S.-J. Rey, \sl Collective Coordinate Quantization of Dirichlet Brane, 
\tt hep-th/9604037 \rm, in the Proceedings `Frontiers in Quantum Field Theory'
Toyonaka 1995, pp. 74 - 85 (World Scientific, Singapore, 1996).}

\nref\pasq
{A. Pasquinucci, \sl On the Scattering of Gravitons on Two Parallel D-Branes,
\tt hep-th/9703066 \rm.}

\nref\factorization
{J. Polchinski, \sl Factorization of Bosonic String Amplitude, 
\rm \np{307}{1988}{61}.}

\nref\oyvind
{V. Periwal and \O. Tafjord, \sl D-brane Recoil, 
\rm \prd{54}{1996}{3690}, \tt hep-th/9603156. \rm}

\nref\fisch
{W. Fischler, S. Paban, M. Rozali, \sl Collective Coordinates for D-branes, 
\rm \pl{381}{1996}{62}, \tt hep-th/9604014. \rm}

\nref\dasrey{
S.R. Das and S.-J. Rey, 
\sl Dilaton Condensates and Loop Effects in Open and Closed Bosonic String, 
\rm \pl{186}{1987}{328}.}

\nref\polchinskicomb{
J. Polchinski, \sl Combinatorics of Boundaries in String Theory \rm, 
\prd{50}{1994}{6041}.}

\nref\lifschytz{
G. Lifschytz, \sl Probing Bound States of D-branes, \tt hep-th/9610125 \rm}

\nref\sanjaye{
Z. Guralnik and S. Ramgoolam, \sl Torons and D-brane Bound States,
\tt hep-th/9702099. \rm}

\nref\gns{
E. Gava, K.S. Narain and M.H. Sarmadi,
\sl On the Bound States of p- and (p+2)-Branes,
\tt hep-th/9704006. \rm}

\nref\angle{
M. Berkooz, M.R. Douglas and R.G. Leigh, \sl Branes Intersecting at Angles,
\rm \np{480}{1996}{265}, \tt hep-th/9606139. \rm}

\nref\vijay{V. Balasubramanian and R.G. Leigh,
\sl D-branes, Moduli and Supersymmetry, \tt hep-th/9611165. \rm}

\nref\akiwati{
A. Hashimoto and W. Taylor,
\sl Fluctuation Spectra of Tilted and Intersecting D-branes from the 
Born-Infeld Action, \tt hep-th/9703217. \rm}

%%%%%%%%%%%%%%%%% Text of the paper %%%%%%%%%%%%%%%%%%%%%%%%%%%%%%

\newsec{Introduction}
Dirichlet branes \polchinski\
in string theory are {\sl quantum BPS solitons} that are
coupled minimally to the Ramond-Ramond (RR) gauge fields. In unraveling 
nonperturbative aspects of string theories, it has been essential to 
understand spectra of D-brane bound states. For example, strongly coupled
Type IIA string and partonic description via M(atrix) theory relies heavily
on the existence of threshold bound state of D zero-branes for arbitrary
charge. Likewise, conifold transitions in Type II strings compactified on
Calabi-Yau threefold assumes no threshold bound state for wrapped two-branes 
and three-branes near the the singularity. (Non)existence of either types of
bound states are extensively studied \sethi.
A more nontrivial class of D-brane bound states are those of non-threshold 
type, for which binding effect has to be properly taken into account. 
The most interesting one of this kind arises in Type IIB string theory.
The ten-dimensional Type IIB string theory exhibits $SL(2,{\bf Z})$ 
S-duality symmetry \schwarz.  
There are two types of rank-two antisymmetric tensor
potentials, one $B^{(NS)}_{MN}$
from NS-NS sector and another $B^{(RR)}_{MN}$ from Ramond-Ramond sector. 
Fundamental string (F-string) and D-string are electric sources that couples 
minimally to the two tensor potentials respectively. Under the $SL(2,{\bf Z})$
S-duality, the pair of antisymmetric tensor potential as well as pair
of the F- and the D-strings transform as doublets. The $SL(2, {\bf Z})$ 
S-duality predicts existence of infinite orbits of $(m,n)$ strings
carrying NS-NS charge $m$ and RR charge $n$, with $m$ and $n$ relatively 
prime \witten.
The $(m,n)$ strings are BPS configurations annihilated by sixteen supercharges
and have a tension
\eqn\tension{T_{\rm (m,n)} = T \sqrt{m^2 + {n^2 \over g_{IIB}^2}}. }
Here, $T \equiv 1/ 2 \pi \alpha'$ is the F-string tension 
and $g_{\rm IIB}$ denotes Type IIB string coupling parameter.
Investigation of the $(m,n)$ string bound states so far has been focused 
mainly on kinematical aspects such as spectra and degeneracy. Many 
interesting dynamical issues such as formation/dissociation of the 
bound states, recoil and radiation and conformal field theory of 
$(m,n)$ string collective coordinates are largely unexplored. 

In this paper, we initiate systematic study of D-brane {\sl dynamics}
and study formation of $(m,n)$ string bound state and subsequent conversion
of the released binding energy into local recoil.
Consider a macroscopic F-string of charge $(m,0)$ approaching a macroscopic 
D-string of charge $(0,n)$ at 
arbitrarily slow relative velocity. Once their wave functions overlap, 
the F-string can fuse inside the D-string worldsheet by converting itself
into homogeneous electric flux. The conversion is nothing but stringy
Higgs mechanism \reyhiggs\ of $B_{MN}^{(NS)}$ field mediated via 
Cremmer-Scherk coupling \cremmerscherk\ present
in the Dirac-Born-Infeld (DBI) worldsheet action of D-string.       
In section 2, we calculate {\sl inclusive} absorption cross section of
F-string by D-string and show that the cross section is of order 
$g_{\rm IIB}$ in the limit $m \gg n$ but of order $g_{\rm IIB}^2$ for
finite $m$, $n$. Once the $(m,n)$ bound state string is formed, binding energy
has to be either released via radiation of closed string states or 
converted into internal excitations via local recoil of the string.
In string perturbation theory, such information is encoded in the 
higher order contributions to the cross section. 
Restricting to the former limit $m \gg n$, in section 3, we argue that the 
dominant higher order corrections come from local recoil of bound state
string rather than radiation of closed string states. 
In section 4, based on the observation of section 3, 
we calculate leading order correction (annulus diagram) to the absorption
cross section. We show explicitly that rigid recoil effect is suppressed
in the kinematical regime considered but internal excitation via 
local recoil deformation is nonvaninshing and provides 
subleading correction to the total absorption cross section.   
In section 5, we conclude with implications of the result to other $p-(p+2)$
D-brane nonthreshold bound states that are related to the $(m,n)$ string
bound state by a series of S- and T-dualities.

%%%%%%%%%%%%%%%%%%%%%%%%%%%%%%%%%%%%%%%%%%%%%%%%%%%%%%%%%%%%%%%%%%%%%%%
\newsec{Inclusive Absorption Cross Section of Macroscopic F-String}
%%%%%%%%%%%%%%%%%%%%%%%%%%%%%%%%%%%%%%%%%%%%%%%%%%%%%%%%%%%%%%%%%%%%%%%
Consider Type IIB string theory compactified on a circle, say, 
$X^9$ direction of radius $R$. $L = 2 \pi R$ is the period in $X^9$
direction. We will take $R \gg \sqrt{\alpha'}$ so that strings wrapped
on it are macroscopic. Arrange F-string with winding number $m$ and
D-string with winding number $n$, $m,n \gg 1$ around the $X^9$ circle.
To calculate absorption cross section of the F-string by D-string
(or vice versa), we bring the F-string adiabatically to the D-string.
For example, we let the F-string approach the D-string with arbitrarily
small velocity, $v \ll 1$. As the details of final states are not of
direct interest to us, we consider {\sl inclusive} absorption 
cross section, hence, {\sl total} bound state formation rate. By unitarity
and optical theorem, the inclusive cross section is related to the 
forward Compton scattering amplitude. The full process can then be
visualized as follows: the winding F-string meets the D-string target,
split by fusing into the D-string worldsheet and then rejoin back to the
winding F-string. The leading order parton diagram associated with this
forward Compton scattering amplitude is then given by two Type IIB 
winding string vertex operators on a disk diagram with Dirichlet boundary 
condition. (A closely related disk amplitude but with Neumann boundary
condition has been considered previously \dai\ in the context
of decaying macroscopic bosonic string.  )

The ground state configuration of a winding F-string comes from the 
massless modes in the original uncompactified theory, hence, 
it is characterized by a polarization vector $\epsilon_\munu$. 
Momentum quantum number of the winding F-string measured in the rest 
frame of the winding D-string is then given by
\eqn\kinb{\eqalign{
& p_R=(E, mR/\Ap, {\bf p}_\perp),\quad p_L=(E, -mR/\Ap, {\bf p}_\perp),
\cr & E^2- {\bf p}_\perp^2=(mR)^2/\Ap^2.
}}
$p_R$ and $p_L$ are the usual right-moving and left-moving momenta which 
appear in the zero-mode part of $X(z, \bar{z})$.
Define two (dimensionless) kinematic invariants
\eqn\st{\eqalign{
&s \equiv \Ap p_{\parallel}^2 = -\Ap E^2= -  [(mR)^2/\Ap + \Ap \, 
{\bf p}_\perp^2] \cr
&t \equiv {\Ap \over 2} p_1 \cdot p_2. 
}}
The invariant $t$ is defined out of incoming ($p_1$) and outgoing 
($p_2$) momenta. Since we are interested in the forward scattering 
limit, we will take $t\rightarrow 0$ in the end. Note also that 
$s\gg 1$ for $m \gg 1$.  

The forward Compton scattering disk amplitude is given by
\eqn\treea{ 
\calA_{D_2} =  
n (2 \pi^2 T_1 L) \Bigg({\kappa\over 2\pi \Ap \sqrt{L}}\Bigg)^2 
\int_{D_2} {d^2 z_1 d^2 z_2 \over V_{\rm CKV}} 
\langle V_{-1}^\dagger (z_1,\bz_1) V_0 (z_2,\bz_2)\rangle_{D_2}.
}
The notation is as follows: $T_1$ is the D-string tension and 
$\kappa$ is the 10-dimensional gravitational coupling. 
The normalization constant $\kappa/ 2\pi\Ap$ is for the massless closed 
string vertex operators. 
The normalization constant for the disk vacuum amplitude, 
$2\pi^2 T_1$, depends on the convention for conformal killing volume. 
We use the infinitesimal form of the standard representation of 
$SL(2, {\bf R})$ on upper half plane, 
$\delta z = \alpha + \beta z + \gamma z^2$, where 
$\alpha$, $\beta$ and $\gamma$ are real. 
We replace three real integrals into integrals over 
$(\alpha, \beta, \gamma)$, compute an appropriate Jacobian and drop the 
$(\alpha, \beta, \gamma)$ integral. 
The factors of $L$ in \treea\ are inserted to account for the 
compact direction volume. 
As such the amplitude is proportional to the Kaluza-Klein 
zero-brane tension and 
the gravitational coupling in compactified 9-dimensional Type II string. 
From the point of view of 9-dimensional Type II string theory,
the process under consideration is interpreted as scattering a 
massive particle of mass $M =m \, R/ \alpha'$ 
off a fixed zero-brane target.

The winding string state is described by vertex operators $V_{-1}$ and $V_0$ 
in -1 and 0 ghost number picture. In ten-dimensional notation they are given 
by
\eqn\vert{\eqalign{
& V_{-1} (z_1,\bz_1)=\epsilon_{\munu}
:V^\mu_{-1}(p_1,z_1)::{\overline V}^\nu_{-1}(p_1, \bz_1):\cr
& V_0 (z_2,\bz_2)=\epsilon_{\munu}
:V^\mu_0(p_2,z_2)::{\overline V}^\nu_0(p_2, \bz_2):,
}}
where
\eqn\supva{\eqalign{
&V_{-1}^\mu(p,z)=e^{-\phi(z)} \psi^\mu(z) \, e^{ip\cdot X(z)} \cr
&V_0^\mu(p,z) = (\partial X^\mu(z)+ip\cdot \psi(z)\psi^\mu(z)) \, 
e^{ip\cdot X(z)}.
}}

When considering scattering off D-p-branes,
it is useful to use {\sl doubling method} \myers \aki 
and replace ${\bar X}^{\mu}(\bz)$ and  ${\bar \psi}^{\mu}(\bz)$ 
by $D^{\mu}_{\nu}X^{\nu}(\bz)$ and $D^{\mu}_{\nu}\psi^{\nu}(\bz)$. 
The tensor $D^{\mu}_{\nu}$ is defined as 
$D^{\mu}_{\nu}\equiv {\rm diag}(1,\cdots,1,-1,\cdots,-1)$, 
where the first $p+1$ entries are 1. 
The bosonic and fermionic propagators are given by
\eqn\treeprop{\eqalign{
&\brac{X^{\mu}(z)X^{\nu}(w)}=-{\Ap\over 2}\eta^{\mu\nu}\log(z-w)\cr
&\brac{\psi^{\mu}(z)\psi^{\nu}(w)}=-{\Ap\over 2}\eta^{\mu\nu}{1\over z-w}
}}
Using the above propagators, it is straightfoward to evaluate the forward
Compton scattering amplitude. Closely related scattering amplitudes have
been studied extensively already \myers \aki. 
Keeping track of normalization factors carefully, we obtain 
\eqn\treeb{\calA_{D_2} = 
{n T_1 \kappa^2\over 8} 
{ \Gamma(t)\Gamma(s)\over \Gamma(1+s+t)}(sa_1-ta_2)}
$a_1$ and $a_2$ are complicated kinematic factors. 
For our purposes, it is enough to note that 
$a_1 = s\cdot Tr(\epsilon\cdot \epsilon^\dagger)$ plus terms that vanish 
in the forward scattering limit, $t \rightarrow 0$. 
Moreover, $Tr(\epsilon\cdot \epsilon^\dagger) =1$ for the 
polarization tensor normalization adopted above.
 
In the limit $s\gg 1$ and $t \goto 0$, we can Taylor expand the amplitude 
in $t$ and use Stirling's formula for $\Gamma(s)$. Recall also that 
$\Gamma(t) \goto 1/t$ as $t\goto 0$. 
In this approximation, the amplitude simplifies to
\eqn\cutz{\calA_{D_2} \approx {n T_1 \kappa^2\over 8} 
\Big(-s\ln{s} + {s\over t}\Big) + {\rm regular}.}

The $t$-channel pole comes from the dilaton and graviton tadpoles 
and is irrelavant. The imaginary part of $\calA$ comes from the branch cut:
\eqn\imag{{\rm Im}(\calA_{D_2})= {\pi n T_1\kappa^2 s\over 8}}.

The optical theorem relates $Im(\calA)$ to the {\sl total absorption} cross section in the following way:
\eqn\crosec{
\sigma_{\rm absorp} = {2 \, {\rm Im} (\calA_{D_2})\over 2E|v_\perp|} = 
{\pi n T_1 \kappa^2\over 8 |p_\perp|} 
\left({( m \, R)^2\over \Ap} + \Ap \, {\bf p}_\perp^2\right),
}
where the denominator takes into account the flux of the F-string 
in the standard relativistic normalization for one particle state. 
Finally, we express \crosec\ in terms of dimensionless string coupling $g_s$: 
\eqn\fine{
\sigma_{\rm absorp} = {4 g_s \pi^7 \Ap^3 n \over |{\bf p}_\perp|} 
\left({(m \, R)^2\over \Ap} + \Ap \, {\bf p}_\perp^2\right),
}
using the following well-known relations \polchinski 
\eqn\tenkap{
T_p\kappa = \sqrt{\pi}(2\pi \sqrt{\Ap})^{3-p},\ \  
\kappa = 8 \pi^{7\over 2} \Ap^2 g_s.
}

A remark is in order. For a fundamental string with a finite length,
$m=$ finite, the forward Compton scattering disk amplitude exhibits
poles associated with one particle intermediate states only in the
complex momentum plane. As such, the absorption cross section vanishes
identically. For a macroscopically long fundamental string $m \gg 1$,  
however, the infinitely many one particle state poles collapse down
densely along a real axis and creates a branch cut effectively.
The inclusive absorption cross section we have calculated is precisely
in this limit. In fact, the fact that the absorption cross section is
${\cal O}(g_s)$ can be understood alternatively from the absolute value
square of transition amplitude mediated by Cremmer-Scherk coupling
between $B_{MN}^{(NS)}$ and DBI gauge field strength $F_{MN}$.

%%%%%%%%%%%%%%%%%%%%%%%%%%%%%%%%%%%%%%%%%%%%%%%%%%%%%%%%%%%%%%%%%%%%%%%%%%%%%
\newsec{Higher Order Corrections and Local Recoil Deformation}
%%%%%%%%%%%%%%%%%%%%%%%%%%%%%%%%%%%%%%%%%%%%%%%%%%%%%%%%%%%%%%%%%%%%%%%%%%%%%
As we have emphasized repeatedly, the leading order absorption cross section 
of the previous section is exact only in the limit $m\rightarrow \infty$. 
However, simple kinematical consideration shows that the binding energy
left out upon formation of non-threshold bound state out of
F-string and D-string scattering state should be converted into other
forms appropriately. From the same kinematical consideration, it is 
straightforward to see that the binding energy is converted either into
internal excitation of bound state $(m,n)$ string 
or release away by radiation of massless modes such as
graviton, dilaton or NS-NS and RR antisymmetric tensor fields. 

In this section, using $SL(2,{\bf Z})$ S-duality of Type IIB string, we argue
that the dominant channel of binding energy release is into internal 
excitation of the bound state $(m,n)$ string. A similar argument in related
context has been given by Verlinde \verlinde\ recently and we
adopt some of his arguments gratefully in out context.  
As in section 2, we consider the limit $m \gg n$ and, moreover, take $n=1$ for 
simplicity. This combination of F- and D-strings is $SL(2,{\bf Z})$ dual to 
a combination of a F-string and $m$ D-strings, hence, we first focus on the 
dynamics of the latter. 

In the large $m$ limit, the dynamics 
of $m$ D-string is described by (1+1)-dimensional ${\cal N} = 8$ 
supersymmetric gauge theory with gauge group $U(m)$. The worldsheet 
Lagrangian is given by
\eqn\sym{
S = \int d^2 x \, {\rm Tr} \Big[ 
{1 \over 4 g^2_{\rm YM}} F^2_{\alpha \beta}
+ (D_\alpha {\bf X}^i)^2 - {g^2_{\rm YM} \over 4} 
[{\bf X}^i, {\bf X}^j]^2 \Big]. }
The ${\bf X}^i\ \ (i = 1, \cdots, 8)$ are eight transverse coordinates
of the D-string in the light-cone gauge. 
The disk amplitude can then be viewed as a Wilson loop average of the 
D-string. At a fixed string coupling, higher order interactions
modifies the worldsheet topology by creating many handles and holes. They
are just fishnets of Feynman diagrams associated with the above $U(m)$
supersymmetric gauge theory. By the `t Hooft power counting in the large
$m$ limit, planar diagrams will dominate, viz. disk diagram with handles 
attached on it are suppressed compared to that with holes only. 
\medskip
\fig{Planar diagram with many holes}{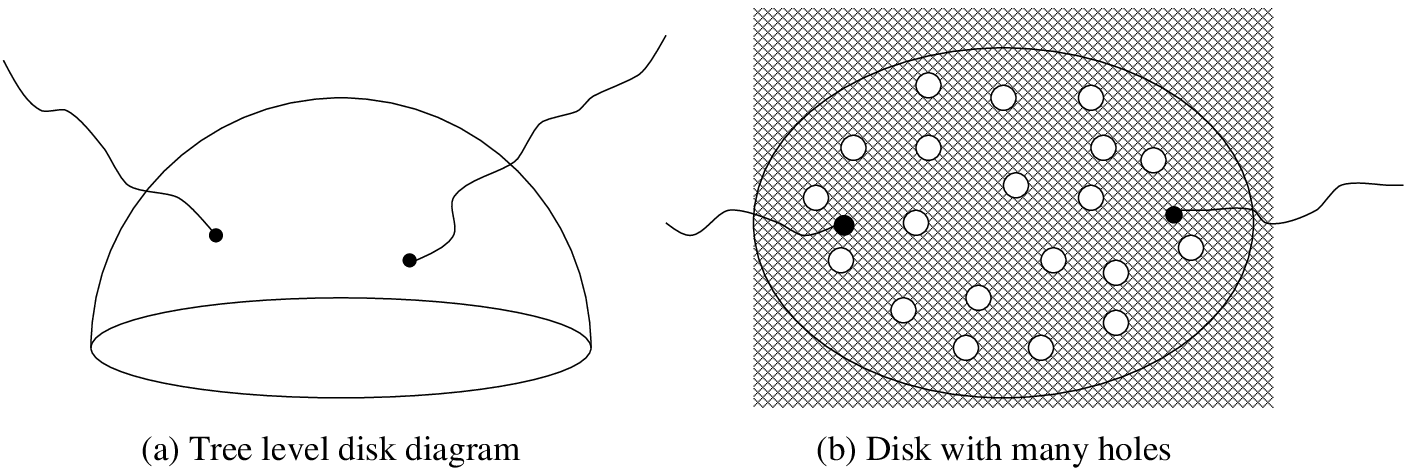}{350pt}
\figlabel\fishnet
\medskip

We now apply $SL(2,{\bf Z})$ S-duality and reinterpret the above
(1+1)-dimensional ${\cal N} = 8$ supersymmetric gauge theory with gauge
group $U(m)$ as a 
M(atrix) theory description of $m$ multiply winding Type IIB F-string. In 
this case, the strong coupling limit of gauge theory corresponds
to weakly coupled Type IIB string theory, the limit we have assumed in
the absorption cross section calculation in section 2. To see this,
consider the large $m$ limit at a fixed string coupling. The bound state
string tension \tension\ is then approximated as\foot{In this section and 
ref. \verlinde, $\Ap$ denotes $2\pi$ times what we usually call $\Ap$}
\eqn\approxtension{
T_{(m,1)} = {1 \over \alpha'}
\sqrt{m^2 + {1 \over g^2_{\rm IIB}}}
\approx {m \over \alpha'} + { (1/ g_{\rm IIB}^2 \alpha') \over 2 m} .}
In the M(atrix) string theory picture, the second term is the energy 
density associated with electric flux $E/L = g^2_{\rm YM} / 2N$ and is
interpreted as contribution of D-string. We thus
obtain the identification
\eqn\coupling{
g^2_{\rm IIB} \alpha' = {1 \over g^2_{\rm YM}  } .}
Note that the D-string tension has been `renormalized' by a factor 
$2m g_{\rm IIB}$. 

The large-$m$ argument for the dominance of planar diagrams, viz. 
disk amplitude in which many holes are nucleated but not handles should 
then hold in the $SL(2,{\bf Z})$ S-dual description as well. Applying this
to the problem we are interested in, 
any higher order corrections in string loop expansion to the forward
Compton scattering disk amplitude are dominated by holes rather than handles
on it. From this observation follows the first of our claim: in the large
$m$ limit the release of binding energy into radiation of Type IIB closed
string massless modes is completely suppressed. What about the planar
diagram contributions?
The boundaries of holes are all with boundary interactions
\eqn\bdryint{
S_{\rm boundary} = \oint d \sigma \, 
[A_\alpha (X^0, X^1) \partial_\sigma X^\alpha + \Phi_i (X^0, X^1)
\partial_\tau {\bf X}^i]. }
They represent gauge field fluctuation on the worldsheet and collective
coordinate fluctuation of the transverse light-cone coordinates 
\reyosaka.
As such, imaginary part of the forward Compton scattering amplitude comes
from various corners of moduli space of the holes. Physically, these
degeneration limit is represented by disk amplitude with lower numbers of
holes and insertion of vertex operators representing gauge field fluctuation
and local recoil of D-string trajectory. This is intuitively transparent.
Overall rigid recoil of the $(m,1)$ string bound state is not possible as is
easily seen from simple kinematical consideration. On the other hand, 
the bound state can support internal excitations in the form of local recoil
of each bound state string bit. The local recoil is quite complicated, hence,
{\sl exclusive} absorption cross section would be technically far more 
involved to calculate. On the other hand, the {\sl inclusive} absorption
cross section we study in this paper is summed over all phase space, hence,
can be straightforwardly calculated from the forward Compton scattering
amplitude. 

From the $SL(2,{\bf Z})$ S-dual point of view, the large $m$ limit 
corresponds to the limit in which the DBI field strength approaches the
{\sl maximal} value $F_{01} = 1/\alpha'$. Physically this means that, 
once the Type IIB string is splitted on the D-string worldsheet, the two
oppositely charged ends are pulled apart very far away from each other. 
In other words, actual Type IIB worldsheet is stretched to a macroscopic
size compared to typical string scale $\alpha'$. In turn, the Type IIB 
string tension is effectively reduced by a large factor, leading to a new
effective tension $T_{\rm eff} = 1/(2m g_{\rm IIB} \alpha')$ ~\verlinde.
  
With the above intuitive understanding of the absorption processs, 
in the next section, we compute the leading order perturbative correction 
to the absorption cross section we have calculated in section 2. 

%%%%%%%%%%%%%%%%%%%%%%%%%%%%%%%%%%%%%%%%%%%%%%%%%%%%%%%%%%%%%%%%%%%%%%%
\newsec{Leading Order Perturbative Corrections}
%%%%%%%%%%%%%%%%%%%%%%%%%%%%%%%%%%%%%%%%%%%%%%%%%%%%%%%%%%%%%%%%%%%%%%%%
In the previous section, we have argued that the worldsheet picture of
fused fundamental string into D-string is a creation of infinitely many
holes, viz. fishnet of planar Feynman diagrams from the M(atrix) gauge
theory point of view. 
In this section, we calculate explicitly the leading order correction
to the absorption cross section in string perturbation expansions and
draw two important conclusions. First, we will show that the rigid recoil
effect is completely suppressed in the kinematical regime under 
consideration. This follows from the fact that logarithmic infrared
divergence arising from exchange of massless Dirichlet open string states
(Bloch-Nordsieck processes) is kinematically suppressed and vanishes
individually. 
Second, we find there are finite corrections of ${\cal O}(g^2_{\rm st})$
to the inclusive absorption cross section. These are effects associated
with local recoil deformation of the F- and D-string bound state.

The next-leading order process in string perturbation expansion is given
by an annulus diagram with insertion of two winding F-string vertex 
operators. We pay particular attention to possible infrared logarithmic
divergences (Bloch-Nordsiek divergences) as well as finite contributions.
Throughout this section we set $\Ap=2$.
\medskip
\fig{The world sheet coordinate for torus and annulus}{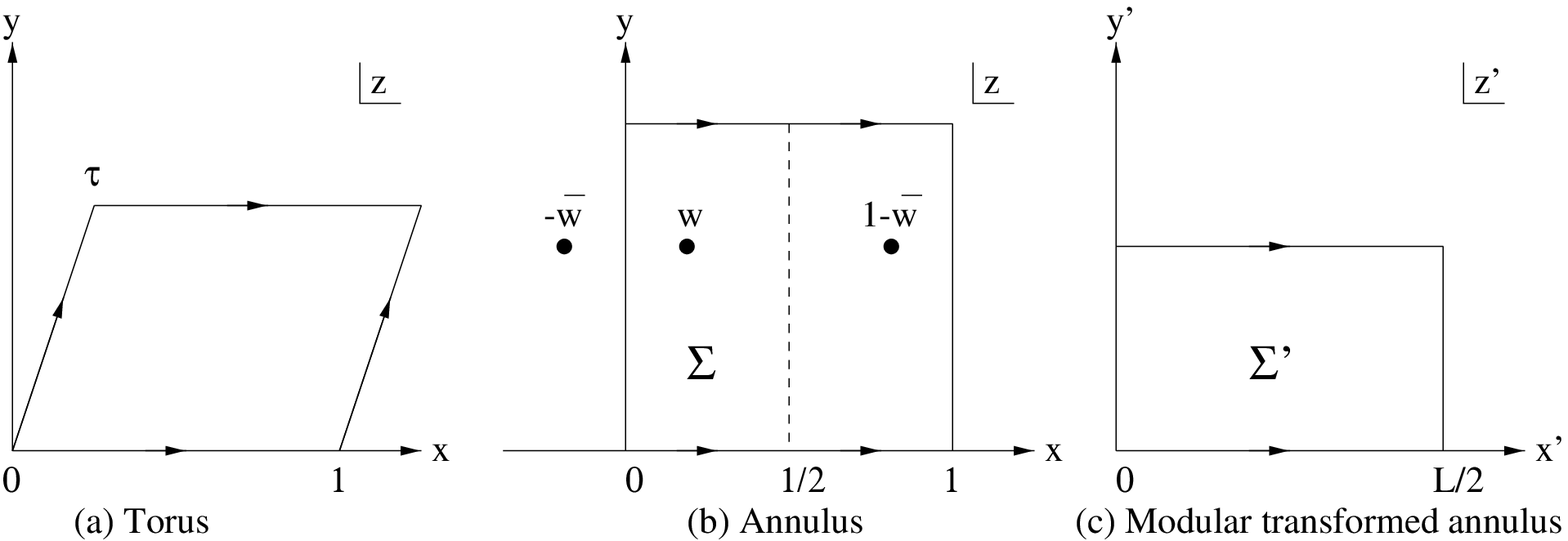}{350pt}
\figlabel\torus
\medskip
%%%%%%%%%%%%%%%%%%%%%%%%%%%%%%%%%%%%%%%%%%%%%%%%%%%%%%%%%%%%%%%%%%%%%%%%
\subsec{Correlators on Annulus}
%%%%%%%%%%%%%%%%%%%%%%%%%%%%%%%%%%%%%%%%%%%%%%%%%%%%%%%%%%%%%%%%%%%%%%%%
The annulus diagram amplitude is conveniently evaluated from ${\bf Z}_2$
involution of torus diagram amplitude. On torus, we introduce standard 
coordinates with a flat metric $z = x + i y$ and complex structure modulus
$\tau$ characterizing the parallelogram. The coordinate system is depicted
in Fig. \torus.
In the following, $z_1=x_1+iy_1$ and $z_2=x_2+iy_2$ will denote the 
positions of the closed string vertices. For notational convenience, 
we will sometimes use $(z,w)$ in place of $(z_1,z_2)$. Similarly, 
we will use $(p,q)$ and $(p_1, p_2)$ interchangebly for the momenta 
of the closed strings.

The propagator on a torus may be obtained by the method of images. 
After regularizing the infinite image charge sum, one finds

\eqn\tprop{
G_T(z,w|\tau) = 
-\log{\left| {\theta_1(z-w|\tau)\over \theta_1'(0|\tau)} \right|^2} 
 + {2 \pi\over \tau_2}( {\rm Im} (z-w))^2.
}
The first term gives rise to the correct short-distance singularity 
($G_T\goto -\log|z-w|^2$ as $z-w\goto 0$) and is periodic in 
$z-w \goto z-w+1$. The second term is needed to ensure periodicity in 
$z-w \goto z-w+\tau$ and flux conservation.

In order to go to an annulus, we introduce a new coordinate $\sigma$, 
such that $z=\sigma_1+\tau \sigma_2$. An annulus is obtained by 
projecting world sheet fields under the mapping 
$\sigma_1 \goto 1- \sigma_1$, $\sigma_2 \goto \sigma_2$. 
Clearly, the distance between two points on the world sheet is invariant 
if and only if $Re(\tau)=0$ and so we can set $\tau = iT$ for a real 
number $T$. 
After the projection, the annulus world sheet ($\Sigma$) is parametrized 
by $z=x+iy$, where $0\le x\le \half$ and $0\le y \le T$. Note that, 
unlike the torus, the annulus does not have modular invariance as 
$\tau \goto -1/\tau$ and $T$ ranges from $0$ to $\infty$.

We can impose the boundary conditions on the propagator again by 
the method of images. Neuman and Dirichlet propagators correspond to 
even and odd projections, respectively, under the mapping mentioned above,

\eqn\aprop{\eqalign{
&G_N(z,w|T)=G_T(z,w|iT)+G_T(z,-\bw|iT)\cr
&G_D(z,w|T)=G_T(z,w|iT)-G_T(z,-\bw|iT).
}}

An image charge at $-\bw$ enforces the boundary condition at $x=0$. 
By the periodicity of $G_T$ on the torus, the same image charge is 
placed at $1-\bw$, which imposes the same boundary condition at $x=\half$. 

A final remark is in order. In the presence of world sheet boundaries, 
the self contraction of $X(z,\bz)$ and $\psi(z,\bz)$ give non-zero 
contributions. In the case of the annulus diagram, the self contraction is
\eqn\self{
G_s(z|T)=
\mp \log\left| {\theta_1(z+\bz|iT)\over \theta_1'(0|iT)} \right|,
}
where minus and plus signs correspond to Neuman and Dirichlet boundary 
condition respectively.

%%%%%%%%%%%%%%%%%%%%%%%%%%%%%%%%%%%%%%%%%%%%%%%%%%%%%%%%%%%%%%%%%%%%%%%
\subsec{Annulus Diagram for Forward Compton Scattering}
%%%%%%%%%%%%%%%%%%%%%%%%%%%%%%%%%%%%%%%%%%%%%%%%%%%%%%%%%%%%%%%%%%%%%%%%
Calculation of the forward Compton scattering on the annulus diagram 
proceeds essentially the same as on the disk diagram. In contracting the
two F-string winding vertex operators, the only role played by the 
correlators of $\partial X$ and $\psi$ is to determine kinematic factors, 
as is explicitly shown, for example, in \pasq.
The integrand of the final expression contain only the zero-mode 
contributions of the partition function and the correlators of 
the exponential factors, $\exp(ik\cdot X)$.
Using the propagators and the self-contraction given in the previous 
sub-section, one obtains
\eqn\expc{\eqalign{
&\brac{ e^{ip\cdot X(z, \bz)} e^{iq\cdot X(w, \bw)} }\cr
&=\exp[-p_\para\cdot q_\para G_N(z,w) -p_\perp\cdot q_\perp G_D(z,w)
- (p_\para^2 - p_\perp^2) G_s(z) - (q_\para^2 - q_\perp^2) G_s(w)]\cr
&= \exp[-t\{G_T(z,w)-G_T(z,-\bw)\}+s\{G_T(z,-\bw)-G_s(z)-G_s(w)\}]\cr
&=\left|{\brac{12}\brac{\bar1\bar2}\over \brac{1\bar2}\brac{\bar1 2}}\right|^t 
\left|{\brac{1\bar1}\brac{2\bar2}\over \brac{1\bar2}\brac{\bar1 2}}\right|^s
\exp\left[-{\pi\over 2T} s(z-\bz-w+\bw)^2\right],
}} 
where we have defined $\brac{ij}=|\theta_1(z_i-z_j|iT)|$ and $\brac{i\bar{j}}=|\theta_1(z_i+\bar{z_j}|iT)|$. 

Using these results, we have found that the forward Compton scattering
amplitude on an annulus diagram is given by
\eqn\loopamp{
\calA_{C_2} 
=iA_0 K\dint{0}{\infty}{dT\over T}T^{-(p+1)/2}\int_{\Sigma}d^4z
\left|{\brac{12}\brac{\bar1\bar2}\over \brac{1\bar2}\brac{\bar1 2}}\right|^t 
\left|{\brac{1\bar1}\brac{2\bar2}\over \brac{1\bar2}\brac{\bar1 2}}\right|^s
e^{{2\pi\over T} s(y_1-y_2)^2},
}
where $A_0$ is a normalization constant defined as
\eqn\ao{
A_0 \equiv n^2 (8\pi^2\Ap)^{-{p+1\over 2}} L 
 \left( {\kappa \over 2\pi  \sqrt{L} } \right)^2  }
and $K=sa_1-ta_2$ is the same kinematic factor as in the disk diagram. 
The measure for the modulus integral is $dT/T$. 
The factor $T^{-(p+1)/2}$ comes from the integration of 
normalized zero-modes in Neuman directions. 
	
Conformal Killing symmetry is modded out by fixing one of the $y$ 
coordinates, dropping the integral and multiplying the integrand by $T$. 
We choose to set $y_1 = 0$ and let $y\equiv y_2$ vary from 
$-T/2$ to $T/2$.

%%%%%%%%%%%%%%%%%%%%%%%%%%%%%%%%%%%%%%%%%%%%%%%%%%%%%%%%%%%%%%%%%%%%%
\subsec{Bloch-Nordsieck Infrared Divergence at $T \goto \infty$}
%%%%%%%%%%%%%%%%%%%%%%%%%%%%%%%%%%%%%%%%%%%%%%%%%%%%%%%%%%%%%%%%%%%%%
The above annulus amplitude \loopamp\ contains an infrared divergence. 
The divergence is due to propagation of massless open string states, viz.
recoil vertex operators in $T \goto \infty$. 
In order to understand the physics behind the divergence, it is 
convenient to perform the follwing conformal mappings

\eqn\thinmap{
\rho=\exp(2\pi i z)\  \, \lambda={\rho-1\over \rho+1}.
}
as depicted in the following figure. 
\medskip
\fig{$T\goto \infty$ limit is a disk with a long, thin strip attached.}{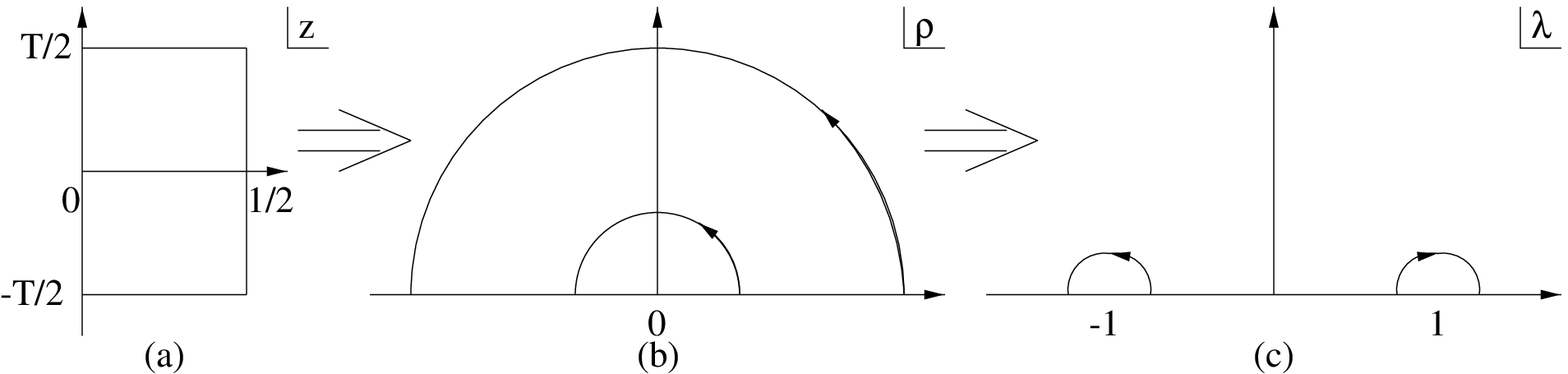}{350pt}
\figlabel\thin
\medskip
The conformal $\lambda$-plane in Fig. \thin (c) is the upper half plane, 
which is equivalent to a disk with two small semicircles removed and 
identified. As $T$ goes to infinity, the semicircles shrink to points on 
the boundary. As discussed above, this is the limit in which 
the annulus diagram factorizes into a disk diagram with Dirichlet 
open string states propagating between the two points on the 
boundary\factorization. 
\medskip
\fig{An annulus diagram and the factorization limit of it.}{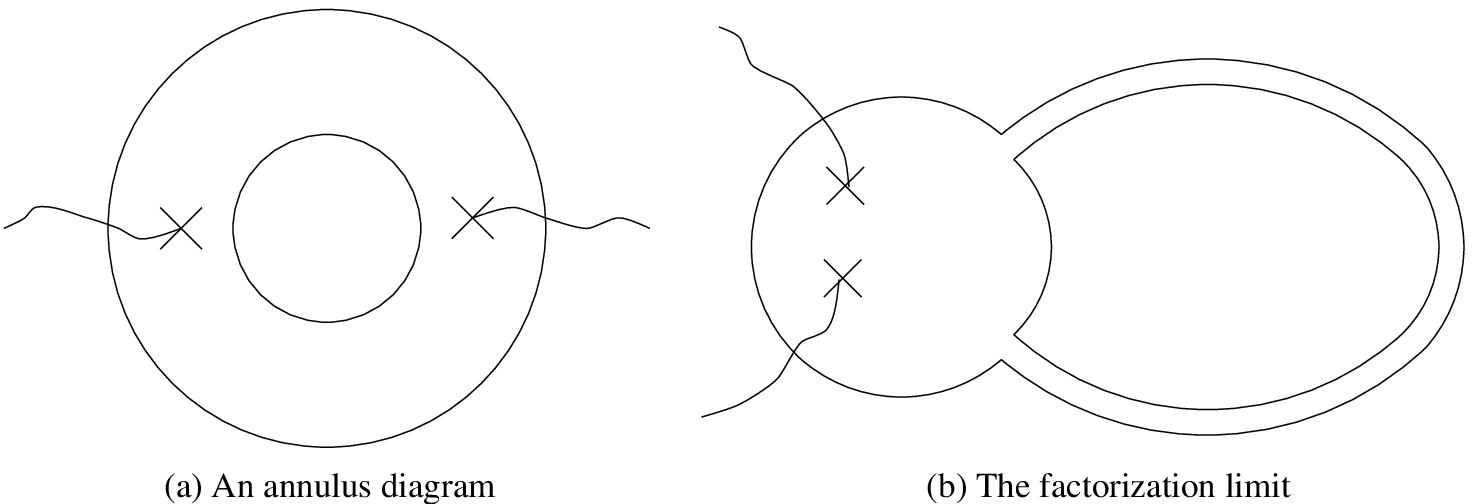}{350pt}
\figlabel\factorization
\medskip
The $T$-variable is the proper time of propagating open string
state and $T$-integral gives rise to a propagator for each open string
states. Clearly, zero momentum limit of the massless open string state
 propagator will give rise to spacetime infrared divergence. The 
zero momentum open string state is nothing but the translational zero
mode of the D-brane represented by the vertex operator $V_o 
= \oint \partial_n {\bf X}^i$. From the
worldsheet point of view, the divergence is due to violation of conformal
invariance due to shrinking annulus open string tadpole.  

The violation of conformal invariance is cured by taking into account of
recoil of the D-string \oyvind \fisch. Recoil of the D-string amounts to 
fluctuation of transverse position of the D-string and is described by 
insertion of recoil vertex operator $V_{\rm recoil} = \oint \Phi_i(X^0, X^1) 
\partial_n {\bf X}^i$ to the disk diagram. The disk diagram with 
recoil vertex operator inserted is not conformally invariant either.
However, when summed over, the two different sources of violation of
conformal invariance cancel each other via Fischler-Susskind mechanism
\dasrey \polchinskicomb \reyosaka. It is in this way energy and 
momentum conservation for scattering process involving D-branes 
is maintained.

Fortunately, in our case, all these complications disappear. This is 
because the inclusive absorption cross section is related to the forward
Compton scattering amplitude. As such, there is no momentum transfer from 
the F-string to the D-string, hence, we expect that the infrared divergence
vanishes in this limit. Below, we will explicitly see that this is 
indeed the case.

It is actually straightforward to separate the potential divergence from 
\loopamp using the following product representation of the theta 
function: 
\eqn\thetarep{
\theta_1(z|iT)=f(T)\sin\pi z 
\prod_{n=1}^{\infty} (1-2e^{-2\pi nT}\cos2\pi z + e^{-4\pi nT}).
}
Since the integrand depends only on the ratio of theta functions, 
the prefactor $f(T)$ is irrelavant. Moreover, for fixed positions
of $z_1$ and $z_2$, the infinite product simply converges to 1. 
Therefore, the leading order divergence of the annulus amplitude takes 
the following simple form:
\eqn\divinf{\eqalign{
&{\rm div} (\calA_{C_2})=
iA_0 K\dint{1}{\infty} {dT\over T} T^{-(p+1)/2}\cdot T\times 4I,\cr
&I \equiv \int dx_1dx_2dy 
\left|{\sin\pi(x_1-x_2+iy)\over \sin\pi(x_1+x_2+iy)} \right|^{2t}
\left[{\sin2\pi x_1 \sin2\pi x_2\over |\sin\pi(x_1+x_2+iy)|^2}\right]^s .
}}
As before, we have kept the Dirichlet boundary directions arbitrary
so that the dependence of divergence on the D-brane dimension $p$ can be
read off. If we cut off the proper-time integral at $T=\Lambda$, we find 
$\sqrt{\Lambda}$ divergence for D-particle ($p=0$), $\log{\Lambda}$ for 
D-string ($p=1$) and no divergence for $p\ge 2$.

%It now remains to perform the $z$ integral in \divinf. 
%In the conformal $\lambda$-plane, the integral becomes
%\eqn\coeffinf{
%I \sim \int {d^2z dy \over |z+1|^2 |z-1|^2 (1+y^2)} 
%\left|{z-iy\over z+iy}\right|^{2t} 
%\left[ {4Im(z)y\over |z+iy|^2}\right]^s,
%}
%where $z$ is coordinate of one of the closed string which is free to move on the whole upper half plane, and $y$ is the position of the other closed string on the $y$-axis. As mentioned above, this expression can be seen 
%as a disk amplitude integrated over all possible propagation of open string states between $-1$ and $1$. Thus we expect to find a manifestly SL(2,R) invariant expression which, upon fixing three real variables, 
%gives \coeffinf\. The result is:
%\eqn\unfix{
%I \sim \int {d^2z d^2w dp dq (p-q)^2 \over |z-p|^2|z-q|^2|w-p|^2|w-q|^2}
%\left|{z-w\over z+\bw}\right|^{2t} \left[ {4Im(z)Im(w)\over |z+\bw|^2}\right]^s
%}
%Fixing the closed strings at $i$ and $iy$ $(0\le y \le 1)$, which is 
%standard in the tree level calculation,
%\eqn\refix{
%I \sim \dint{0}{1}dy\left[{(1-y)^2\over(1+y)^2}\right]^t
%\left[{4y\over (1+y)^2}\right]^s (1-y^2)
%\int {(p-q)^2 dpdq\over (y^2+p^2)(y^2+q^2)(1+p^2)(1+q^2)}.
%}
%The integrand in the $y$ integral is almost identical to the tree 
%calculation. What is better, we can do the $p$ and $q$ integrals
% exactly. %%

After straightforward calculation, one finds
\eqn\infans{I \sim t {\Gamma(t)\Gamma(s)\over \Gamma(1+s+t)}.}
We conclude that the coefficient of the divergence is the tree level 
amplitude times $t$, the momentum-squared transferred to the D-string.
Since we are taking $t \rightarrow 0$ in the end, we conclude that
there is no violation of conformal invariance, hence, no recoil of
the D-string.

%%%%%%%%%%%%%%%%%%%%%%%%%%%%%%%%%%%%%%%%%%%%%%%%%%%%%%%%%%%%%%%%%%%%%%%
\subsec{Finite Corrections from Local Recoil Deformation}
%%%%%%%%%%%%%%%%%%%%%%%%%%%%%%%%%%%%%%%%%%%%%%%%%%%%%%%%%%%%%%%%%%%%%%%
Having shown that there is no infinities associated with rigid recoil of
the D-string, it now remains to calculate finite corrections.
In this subsection, we will find finite corrections arising from the 
$T\ge 1$ region. 
The $T\le 1$ region corresponds to a closed string emitted and re-absorbed 
by the D-brane and is not relevant to our problem. 
We will continue neglecting the infinite product in the theta functions, 
which amounts to neglecting all massive open string modes. 

In the previous sub-section, the analysis was simplified because 
(1) the integration became an infinite strip, which was mapped to the 
upper half plane and (2) the integrand of the $z$-integral converged. 
The finite corrections then come from (1) finiteness of the 
integration region and (2) the deviation of the integrand from its limiting 
value. In the following, we compute these two contributions. It turns 
out that both of them give same result up to an overall constant and
correspond to two open string branch cut.

\medskip
{\sl Correction due to the finite length of the strip.}
\medskip 
It is again easy to work in the upper half plane (Fig. \thin (c)). 
Clearly, the integral \divinf\ over the two semicircles will give 
the correction. Since $T\ge 1$, $e^{-\pi T}$ is always small and 
we can use the approximation $\lambda = i\tan{z_2} = \pm 1 + r e^{i\theta}$, 
where $0\le r \le e^{-\pi T}$ and $0\le \theta \le \pi$. 
The other closed string vertex is fixed on the imaginary axis 
with coordinate $b = \tan{x_1}$.
Note also that the integrals over the two semicircles yield 
the same value. We have, to the leading order in $e^{-\pi T}$,
\eqn\cutaa{2\cdot\half \int {rdrd\theta db \over r^2(1+b^2)}
\left|{1-ib\over 1+ib}\right|^{2t} 
\left[ {4r\sin\theta\cdot b \over |1+ib|^2} \right]^s.
}
After some algebra, we obtain
\eqn\cutbb{{s\over 32\pi^2} e^{-\pi sT}
\left[{\Gamma(s)\over \Gamma(1+s/2)^2}\right]^2.
}
 
This particular combination of $\Gamma$-functions in the square bracket 
is exactly the one found in the tree level amplitude for a closed string 
absorbed by the D-string and to excite two massless open string modes \aki. 
Thus we recognize this correction as coming from two open string branch
cut. Integration over $T$ produces a numerical constant, which we do not 
compute.

\medskip
{\sl Correction due to deviation of integrand from limiting value}
\medskip
Rescale the $y$ coordinate in \divinf\ by $y \goto yT/2$ so that 
$y$ ranges between $-1$ and $+1$. The integrand becomes
\eqn\cuta{
T\times \left[{\cosh\pi Ty - \cos2\pi(x_1-x_2)\over 
\cosh\pi Ty - \cos2\pi(x_1+x_2)}\right]^t
\left[{2\sin2\pi x_1 \sin2\pi x_2 e^{\half \pi T y^2}
\over \cosh\pi Ty - \cos2\pi(x_1+x_2)}\right]^s.
}
For a fixed value of $y$, the integrand converges to $0$ as $T 
\rightarrow \infty$ and non-zero contribution comes from a domain 
$|y|\le 1/T$. 
For a large value of T, we have
\eqn\cutb{
\left[4\sin2\pi x_1 \sin2\pi x_2 e^{-\half \pi T(2y-y^2)}\right].
}

The factorized $x_{1,2}$ interals yields
\eqn\cutc{
\Big( \dint{0}{1/2}(2\sin2\pi x)^s dx \Big)^2 = \Big[
\half s{\Gamma(s)\over \Gamma(1+s/2)} \Big]^2,
}
the combination that appears in the tree level amplitude of a closed 
string absorbed by the D-string and to excite two massless open string
modes. This is again the two open string branch cut.
The remaining $y$ and $T$ integrals give rise to a numerical factor,
which we again do not compute.

We conclude that the leading order finite corrections come from 
exchange of two open string massless states viz. gauge fields and
translation zero-modes. The correction corresponds to 
local recoil deformation of the F- and D-string bound state.

\newsec{Discussions}
%%%%%%%%%%%%%%%%%%%%%%%%%%%%%%%%%%%%%%%%%%%%%%%%%%%%%%%%%%%%%%%%%%%%%%%%%%%%%%
In this paper, we have studied dynamics of bound state formation between
fundamental and Dirichlet strings in Type IIB string theory. 
We have calculated total inclusive absorption cross section via optical
theorem from the forward Compton scattering amplitude of an F-string off
a D-string target. 
We have found that the leading order disk diagram gives cross section
of order ${\cal O}(g_s)$ and
agrees with power counting from stringy Higgs mechanism via 
Cremmer-Scherk coupling. Using M(atrix) string theory and SL(2,${\bf Z})$
mapping, we have argued that higher order corrections come from disk 
diagram with arbitrarily large number of holes and describes conversion
of binding energy into local recoil deformation. To check this, we have 
calculated explicitly leading order corrections from annulus diagram
and have found that rigid recoil of the bound state is absent and that 
two open string state branch cut is present as expected.
 
By a series of S- and T-dualities, one can relate the $(m,n)$ bound-state
string to other configurations of strings and branes. For example, 
T-duality 
in $X^7$ and $X^8$ directions followed by S-duality turn the $(m,n)$ 
string into $m$ D-strings and $n$ D-3-branes, which is T-dual to 
any $p-(p+2)$ bound states \lifschytz \sanjaye \gns. 
Additional T-duality in a direction not parallel to coordinate axis 
give rise to branes intersecting at angles \angle \vijay. 
Early studies of these brane configurations have been focused on their
mass spectrum, degeneracy and supersymmetry. Recently, there have been 
some progress in analyzing their dynamics by using scattering of a 
D-0-brane \lifschytz\ and low energy Born-Infeld action 
\sanjaye \akiwati \gns. 
One notable progress was made in Ref. \gns, where it was shown that the 
condensate of tachyonic modes of ND strings connecting $p-$ and $(p+2)-$branes 
account for the binding energy predicted by the BPS formula. 
In spite of all the progress, the details of the {\sl dynamics} of 
non-threshold brane bound states, brane-anti-brane configuration and 
intersecting branes are still poorly understood. We hope that our analysis 
will be useful in future investigations on these issues.

\bigskip
\centerline{\bf Acknowledgements}

We thank C.G. Callan, %J. Castelino, 
E. D'Hoker, %A. Guijosa, 
A. Hashimoto, \O. Tafjord and H. Verlinde for useful discussions. 
The work of S.L. was supported in part by DOE grant DE-FG02-91ER40671.
The work of S.-J.R. was supported in part by the NSF-KOSEF Bilateral
Grant, KOSEF Purpose-Oriented Research Grant 94-1400-04-01-3 and
SRC Program, Ministry of Education Grant BSRI 97-2410, the Monell
Fundation and the Seoam Foundation Fellowships. 

\listrefs

\end